\documentclass[aps,pra,english,twocolumn,letterpaper,notitlepage]{revtex4-1}

\usepackage[T1]{fontenc}
\usepackage{graphicx}
\usepackage{epstopdf}
\usepackage{amssymb}
\usepackage{amsmath}
\usepackage[urlcolor=blue,hyperindex,colorlinks,bookmarks=true,linkcolor=red,citecolor=black]{hyperref}
\usepackage[normalem]{ulem}
\usepackage[abs]{overpic}
\usepackage{color}
\usepackage{rotating}
\usepackage{physics}
\usepackage{bm}
\usepackage{soul}

\usepackage[english]{babel}
\usepackage{blindtext,tikz}
\usetikzlibrary{calc}

\begin{document}

\title{Quantum reservoir computing with a single nonlinear oscillator}

\author{L.~C.~G.~Govia}
\email{luke.c.govia@raytheon.com}
\author{G.~J.~Ribeill}
\author{G.~E.~Rowlands}
\author{H.~K.~Krovi}
\author{T.~A.~Ohki}
\affiliation{Raytheon BBN Technologies, 10 Moulton St., Cambridge, MA 02138, USA}

\begin{abstract}
  Realizing the promise of quantum information processing remains a daunting task, given the omnipresence of noise and error. Adapting noise-resilient classical computing modalities to quantum mechanics may be a viable path towards near-term applications in the noisy intermediate-scale quantum era. Here, we propose continuous variable quantum reservoir computing in a single nonlinear oscillator. Through numerical simulation of our model we demonstrate quantum-classical performance improvement, and identify its likely source: the nonlinearity of quantum measurement. Beyond quantum reservoir computing, this result may impact the interpretation of results across quantum machine learning. We study how the performance of our quantum reservoir depends on Hilbert space dimension, how it is impacted by injected noise, and briefly comment on its experimental implementation. Our results show that quantum reservoir computing in a single nonlinear oscillator is an attractive modality for quantum computing on near-term hardware.
\end{abstract}

\maketitle

\section{Introduction}

Over the last several decades, quantum information science has emerged as a transformative framework for information processing, from high-performance computing, to communication and cryptography \cite{Nielsen10}. Despite the tremendous potential, only recently has a computational task that is prohibitively difficult on a classical computer been demonstrated on quantum hardware \cite{Arute:2019aa}. Intrinsic noise and spurious error remain the roadblock to widespread quantum advantage, and the full power of quantum information processing awaits the demonstration of logical error correction and fault-tolerance \cite{RevModPhys.87.307,Campbell:2017aa}. In the near-term, while we remain in the noisy intermediate-scale quantum (NISQ) era \cite{Preskill:2018aa}, open questions remain as to what, if any, quantum advantage can be expected from current technologies, and how to design noise-resilient computational methodologies that best take advantage of the quantum resources available \cite{Riste:2017}.

In the high-performance computing frontier, neuromorphic, or brain-inspired, computing modalities \cite{Schuman17} show considerable promise, with the canonical example the wide-spread success of neural network approaches to machine learning. A particularly attractive neuromorphic computing approach is reservoir computing \cite{Maass:2002aa,Jaeger:2004aa,Verstraeten:2007aa}, which harnesses the computational power of a disordered, densely-connected, nonlinear network. The network connections are untrained and remain fixed, which both drastically reduces the cost of training, and removes the susceptibility to error in the assignment of flexible network connections, as in traditional neural networks. Recently, this approach to neuromorphic computing has been expanded to include quantum mechanical systems as the reservoir, defining so called quantum reservoir computing \cite{Fujii:2017aa,Ghosh:2019aa}. Theoretical studies have shown application to problems in classical computing \cite{Kutvonen:2018aa,Schuld:2019aa,Chen:2019aa,Nakajima:2019aa,Wright:2019aa} as well as quantum computing \cite{Ghosh:2019ab,Ghosh:2020aa}, and first experimental demonstrations have been reported \cite{Negoro:2018aa,Chen::2020,Dasgupta:2020}.

In this work, we present quantum reservoir computing (QRC) in a continuous-variable system, with a reservoir formed by a single nonlinear oscillator, and contrast to classical reservoir computing (CRC) with the equivalent classical reservoir. By using a continuous-variable system, we reduce the costly repetitions necessary to obtain accurate measurement of expectation values, an issue that affects the run time of discrete-variable quantum machine learning approaches \cite{Wright:2019aa}. We expect this to be an advantage that continuous-variable reservoirs will show over discrete-variable ones in practical implementation. We demonstrate via numerical simulation an improvement in performance of QRC compared to CRC for the same classical task of sine phase estimation. This improvement is both in average error for small training set sizes, and a reduction in performance spread across reservoir parameters. We isolate the main cause of this quantum improvement, presenting compelling evidence that it is due to the intrinsic nonlinearity of quantum measurement.

This paper is organized as follows. In section \ref{sec:RC} we briefly review reservoir computing, and in section \ref{sec:NORC} we introduce quantum and classical reservoir computing with a single nonlinear oscillator. We describe our chosen task in section \ref{sec:task}, and present the results of our numerical simulations in section \ref{sec:Results}, studying the quantum improvement and its origin, as well as the effects of Hilbert space dimension and injected noise. Finally, we present our concluding remarks and a discussion of experimental implementations in section \ref{sec:conc}.

\section{Reservoir Computing}
\label{sec:RC}

The simplest implementation of a reservoir computer consists of a network of $N$ densely connected nonlinear nodes, into which data is fed in a time series manner. If the $N\times1$ vector $\vec{u}(t)$ defines the input signal at time $t$, then the input to the reservoir nodes is given by $\mathbf{W}_{\rm in}\vec{u}(t)$, where $\mathbf{W}_{\rm in}$ is the $N\times N$ input weight matrix. In practice, $\mathbf{W}_{\rm in}$ is fixed for a given task, and $N$ can be replaced by $N_{\rm in}$ if data is input to only a restricted number of reservoir nodes.

Without loss of generality we are free to assume each node in the reservoir has a single degree of freedom, and the full state of the reservoir at time $t$ is described by the vector $x(t)$. The time evolution of the reservoir is given by the nonlinear differential equation
\begin{align}
  \partial_t\vec{x} = f\left(\mathbf{W}_{\rm res}\vec{x}(t),\mathbf{W}_{\rm in}\vec{u}(t),\mathbf{W}_{\rm fb}\vec{x}(t-\tau)\right),
\end{align}
where $f(\vec{a},\vec{b},\vec{c})$ is a nonlinear function. Its first argument describes the internal interaction between reservoir nodes with corresponding internal weight matrix $\mathbf{W}_{\rm res}$. Its second describes the input signal to the reservoir, as previously discussed. The final input to $f$ describes time-delay feedback in the reservoir interaction, with weight matrix $\mathbf{W}_{\rm fb}$. In this work we will not consider any explicit time-delay feedback, and rely on the internal reservoir dynamics to serve as short-term ``memory''.

We label the output signal of the reservoir as $\vec{s}(t)$, which in many cases is simply the internal state $\vec{x}(t)$. In order to compute with the reservoir, we discretize the time-dependent output signals of all $N$ reservoir nodes into $K$ time steps, and collect them into an $N \times K$ matrix $\mathbf{s}_t$. For our purposes we column stack this to form a $NK \times 1$ vector $\vec{s}_t$. The final step in a reservoir computing architecture is application of the $L\times NK$ output weight matrix to the output signal, to obtain the computed output
\begin{align}
  \vec{y} = \mathbf{W}_{\rm out}\vec{s}(t), \label{eqn:ResOut}
\end{align}
where $y$ is the $L\times 1$ task output vector, and (ideally) is the answer to the problem the reservoir computer aims to solve. As with the input weight matrix, if the output signal is only recorded from a subset of nodes $N_{\rm out}$, then the output weight matrix can be reduced to $L\times (N_{\rm out}K)$ in size.

In reservoir computing, only the values of the output weight matrix are trained in a supervised learning fashion. We use the standard approach for training
\begin{align}
  \mathbf{W}_{\rm out} = \mathbf{Y}\mathbf{S}_t^\intercal\left(\mathbf{S}_t\mathbf{S}_t^\intercal + \gamma\mathbb{I}\right)^{-1}, \label{eqn:train}
\end{align}
where $\mathbf{Y}$ and $\mathbf{S}_t$ are $L\times M$ and $(N_{\rm out}K) \times M$ matrices that contain the data for $M$ training instances of the task. $\mathbf{Y}$ contains the correct output of the task, and $\mathbf{S}_t$ the reservoir output signals $\vec{s}_t$, for each training instance. Here $\gamma$ is a ridge-regression parameter used to prevent overfitting.

As the training phase for a reservoir computer requires only a single matrix inversion, it offers considerable computational savings over traditional neural networks. Note that the way we have designed our training procedure allows it to access both correlations in the output signal between nodes at a given instance of time, and across time steps, which stems from the fact that we train using the full output signal $\vec{s}_t$. We find that this gives the best performance. To further reduce the computational cost of training, one can introduce a block-diagonal $\mathbf{W}_{\rm out}$ that does not mix reservoir output signals at different time steps.

\section{A Single Nonlinear Oscillator as a Reservoir}
\label{sec:NORC}

An appealing aspect of reservoir computing is that in principle any nonlinear system can be used as the reservoir. For the model of quantum reservoir computing we consider here, the input and output from the reservoir will be a classical data stream, and the reservoir consists of a single nonlinear oscillator. While it might appear that a single oscillator is a reservoir with only one node, this is in fact not the case. In our formalism, the number of nodes in the reservoir map to the number of independent degrees of freedom in the oscillator.

In a classical nonlinear oscillator, there are two independent degrees of freedom; we choose these to be the position, $X$, and momentum, $P$, quadratures. In a quantum nonlinear oscillator, a state $\rho$ is typically described in the eigenbasis of the number operator, the Fock states. We will instead describe states by a specification of expectation values of the form
\begin{align}
  E_{nm} = {\rm Tr}\left[\rho\hat{X}^n\hat{P}^m\right], \label{eqn:Enm}
\end{align}
for integers $n,m \geq 0$, where
\begin{align}
  \hat{X} = \frac{1}{\sqrt{2}}\left(\hat{a} + \hat{a}^\dagger\right),~\hat{P} = \frac{-i}{\sqrt{2}}\left(\hat{a} - \hat{a}^\dagger\right),
\end{align}
are the usual canonical quadratures, with $\hat{a}$ the lowering operator. Each $E_{nm}$ is an independent parameter, so given the infinite dimension of a quantum oscillator's Hilbert space, it is tempting to assume that such a quantum reservoir is infinite in size.

However, for many realistic states of the oscillator the total number of independent $E_{nm}$ is finite. An extreme example is Gaussian states, which are fully described by the set of expectation values with $n + m \leq 2$, but even an input power restriction will set a limit to the largest occupied Fock state, which in turn implies only a finite number of independent $E_{nm}$. Nevertheless, a quantum nonlinear oscillator can have more degrees of freedom than its classical analogue, and the exponential growth of Hilbert space for oscillators is analogous to the arguments used to motivate the power of quantum reservoirs built from qubit networks \cite{Fujii:2017aa}.

Throughout this work, we consider a Kerr nonlinear oscillator as our reservoir. The quantum version of this system is described by the Hamiltonian (in a frame rotating at the oscillators' frequency)
\begin{align}
  \hat{H}(t) = K\hat{a}^\dagger\hat{a}\hat{a}^\dagger\hat{a} + \alpha u(t)\left(\hat{a} + \hat{a}^\dagger\right)
\end{align}
where $K$ is the Kerr nonlinearity, and $\alpha$ is an overall amplitude to the scalar input $u(t)$. As can be seen, we choose a linear input coupling which drives a single degree of freedom of the reservoir, namely the $P$-quadrature. As output we measure the expectation value of the $X$-quadrature (again only one degree of freedom)
\begin{align}
  s(t) = E_{10} = {\rm Tr}\left[\rho(t)\hat{X}\right],
\end{align}
and we always start the oscillator in a deterministic state, in this case vacuum.

The evolution of the quantum reservoir is given by the Lindblad master equation \cite{Breuer:2002aa}
\begin{align}
  \dot{\rho} = -i\left[\hat{H}(t),\rho\right] + \kappa\mathcal{D}[\hat{a}]\rho, \label{eqn:QRmodel}
\end{align}
where $\mathcal{D}[\hat{x}]\rho = \hat{x}\rho\hat{x}^\dagger - \left\{\hat{x}^\dagger\hat{x},\rho\right\}/2$ is the usual dissipator that describes evolution due to interaction with the environment; in this case, decay of oscillator photons into the environment at a rate $\kappa$. Considering open system evolution is important, both to ensure our model is realistic, and as it can reduce susceptibility to over-fitting, thereby improving performance.

As a model of the classical reservoir, we use the equation of motion for a single nonlinear oscillator given by
\begin{align}
  \dot{a} = -iK(a - 2a^*) - \frac{\kappa}{2}a -i\alpha u(t), \label{eqn:CRmodel}
\end{align}
where $a = (X + iP)/\sqrt{2}$ is the complex scalar amplitude of the classical oscillator, with $a^*$ its complex conjugate. Note that for the quantum model of Eq.~\eqref{eqn:QRmodel}, due to the non-commuting nature of the $\hat{X}$ and $\hat{P}$ operators, and the Kerr nonlinearity, there are many classical models that can be derived from it. We have found that the choice of the specific classical model does not qualitatively affect the reservoirs' performances, and the classical model we have chosen gives good dynamical agreement with the quantum model in parameter regimes where this is expected ($K \ll \alpha, \kappa$).

\section{Sine Wave Phase Estimation}
\label{sec:task}

The task we consider for our reservoirs is the estimation of the phase of an oscillatory signal. This fundamental task in signal processing is challenging as it embodies a nonlinear, non-convex optimization problem \cite{Kay1993}. Explicitly, the input to the reservoir is
\begin{align}
  u(t) = \alpha\sin(\omega_{u} t + \phi),
\end{align}
for a fixed, known frequency $\omega_u$, and an unknown phase $\phi$ in the interval $[0,\pi/2]$ \footnote{We found performance decreased significantly if we increased the interval size. We attribute this to the reservoir confusing phases in $[0,\pi/2]$ with those in $[\pi/2,\pi]$ for which $\abs{\sin(\phi)}$ is the same. This may be related to the parity of the reservoir nonlinearity, and is a topic of future study.}. The task is to estimate the numerical value of this unknown phase using the reservoir computer. As such, the output weight ``matrix'' is a $1 \times N_{\rm out}K$ vector, and the reservoir phase estimate is given by
\begin{align}
  \phi^{\rm est} = \mathbf{W}_{\rm out}.\vec{s}_t.
\end{align}
The optimal weight matrix is determined by training on a set of known phases, and throughout this work we will use a set of $M$ equidistantly spaced phases in the interval $[0,\pi/2]$ as our training set.

We measure the performance of the reservoir at the sine wave phase estimation task via the root mean square (RMS) error of the estimated phases for a test set of size $T$, given by
\begin{align}
  r = \sqrt{\frac{1}{T}\sum_{j=0}^T\abs{\phi_j^{\rm est} - \phi_j^{\rm act}}^2},
\end{align}
where $\phi_j^{\rm act}$ is the actual value of the phase for the $j$'th element of the test set. The test set consists of uniformly random phases from the interval $[0,\pi/2]$.

\section{Results}
\label{sec:Results}

We compare the performance of quantum and classical reservoirs with Gaussian distributed parameters $K$, $\kappa$, $\alpha$, and $\omega_u$, with average values $\bar{K}/\bar{\kappa} = -2$, $\bar{\omega}_u/\bar{\kappa} = 10$, $\bar{\alpha}/\bar{\kappa} = 6$, and $\bar{\kappa} = 1$ in arbitrary units. Each parameter's Gaussian distribution has a standard deviation that is $10\%$ of its average value. Simulation of the reservoir evolution, Eqs.~\eqref{eqn:QRmodel} and \eqref{eqn:CRmodel}, is done in {\tt Julia} \cite{Bezanson:2017aa}, and the quantum reservoir simulation uses the master equation solver package {\tt MESolve.jl} \cite{MESolve}. Each input signal lasts for a time duration of $2/\bar{\kappa}$, and we measure the output at a uniformly distributed set of $K = 100$ points across this time, i.e.~an output sample rate of $50\bar{\kappa}$.

For all the results presented in the following, training is done on an equidistantly spaced grid of $M$ training phases on the interval $[0,\pi/2]$. We find this improves performance compared to training with uniformly random phases from $[0,\pi/2]$, as the equidistant grid ensures that the reservoir is trained with phases that span the full interval and are representative of all possible inputs. Further, we do not fix the ridge regression parameter of Eq.~\eqref{eqn:train}, but find an optimal value in each unique case (quantum vs.~classical, training size $M$, Hilbert space dimension, etc.) by searching over values from $\gamma = 10^{-12}$ to $\gamma = 1$ in powers of ten, and report the smallest value of the RMS error found.

\subsection{Quantum-Classical Improvement}
\begin{figure}[!t]
  \includegraphics[width=0.95\columnwidth]{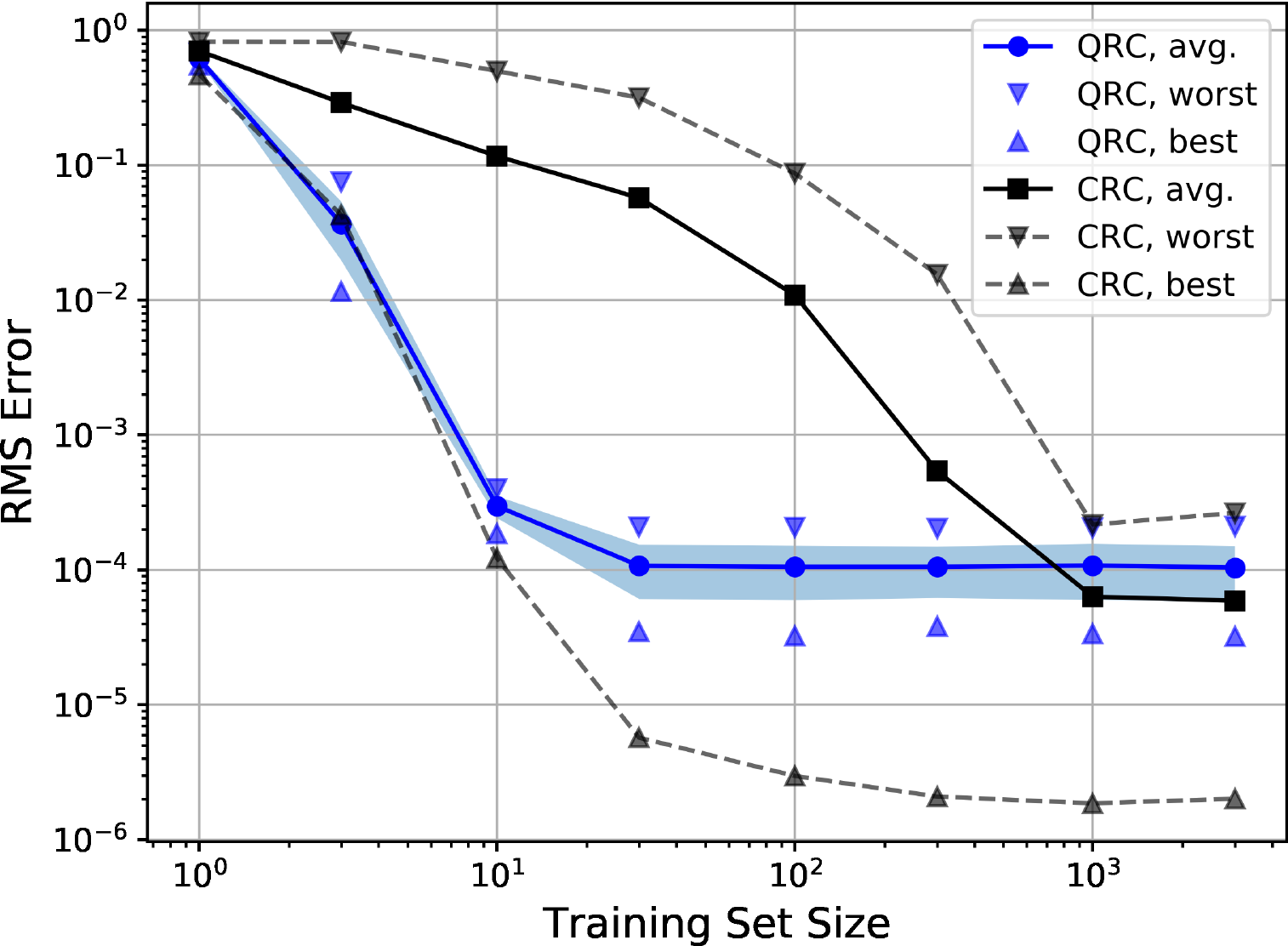}
  \caption{{\bf Quantum and classical reservoir performance.} As a function of training set size, average root mean square (RMS) error on the test set for the quantum nonlinear oscillator reservoirs (QRC) with Hilbert space dimension $d=12$, and the classical nonlinear oscillator reservoirs (CRC). The average results are for 101 reservoirs with Gaussian distributed parameters around the mean values described in the main text. The reservoir output is sampled at 100 equally spaced points in time, and the test set size is 5000. One standard deviation for the QRC performance is shown in the shaded blue region. The RMS error for the QRC and CRC with best and worst case performance (determined independently for each training set size) is also shown.}
  \label{fig:QRCvCRC}
\end{figure}

We first compare the performance of the quantum and classical reservoirs as a function of training set size, as shown in Fig.~\ref{fig:QRCvCRC}, with a Hilbert space dimension of $d=12$ used for the QRC simulations. For smaller training set sizes, especially at intermediate scale around 30, there is a multiple order of magnitude smaller average RMS error for the quantum reservoirs compared to the classical reservoirs. This is the first example of an improvement for QRC compared to the equivalent CRC model. This improvement exists for specific QRC/CRC parameter sets, and is true on average, but is not a universal property for all parameter values. There are parameter sets ($K$, $\omega$, $\alpha$, $\kappa$) where CRC outperforms QRC, as is shown in Fig.~\ref{fig:QRCvCRC} by the best case performance curves for CRC and QRC. Each point on these curves is the lowest RMS error for the RC model, and this best case performance does not occur for the same parameters at each training set size. Nevertheless, there is a demonstrable improvement for QRC over CRC for many specific parameter sets and on average, even if there is no improvement for QRC over all CRC in nonlinear oscillators.

The second important point highlighted in Fig.~\ref{fig:QRCvCRC} is the performance spread of the QRC and CRC models, which we quantify by the RMS error spread factor
\begin{align}
  {\rm RMSE~Spread~Factor} = \frac{{\rm Best~case} - {\rm Worst~case}}{{\rm Average}}. \label{eqn:spread}
\end{align}
The QRC spread is at worst less than a factor of 2, as highlighted by the narrow shadowed region representing one standard deviation in RMS error. By comparison, the CRC spread is typically around a factor of 10 and can be as much as a factor of 30.

This demonstrates the second improvement of QRC over CRC nonlinear oscillators at sine phase estimation: reliability. As we have shown, the performance of the quantum reservoirs is not highly dependent on the specific oscillator parameters, while the performance of a given classical reservoir cannot be inferred from the performance of a different classical reservoir, even one with very similar oscillator parameters. This is important in simulation and design, where the QRC requires far less parameter optimization to find an effective reservoir, as well as in practical implementation, which will naturally have a spread in parameters due to errors in device fabrication, control, and measurement.

\begin{figure*}[!ht]
  \includegraphics[width=2\columnwidth]{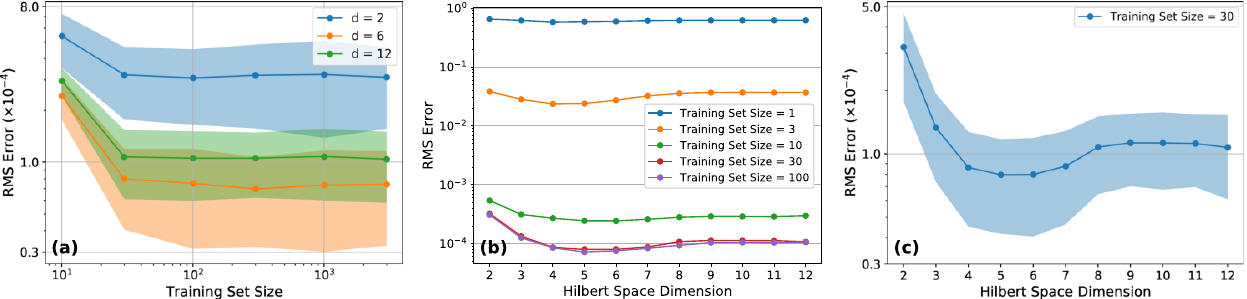}
  \caption{{\bf Hilbert space dimension effect on QRC performance.} (a) Average RMS error as a function of training set size for quantum nonlinear oscillator reservoirs with various Hilbert space dimension. One standard deviation for 101 reservoirs with Gaussian distributed parameters is shown in the shaded regions. (b) and (c) Average RMS error as a function of Hilbert space dimension, for (b) various training set sizes, and (c) for a training set size of 30, which is near optimal. In (c) one standard deviation is shown in the shaded area.}
  \label{fig:dimsweep}
\end{figure*}

\subsection{Hilbert Space Dimension Dependence}

To elucidate the source of the quantum-classical improvements, we begin by studying how the performance of the quantum reservoirs depends on the Hilbert space dimension of the simulations. The results shown in the previous subsection are for a dimension chosen such that the simulations are a good approximation to the infinite dimensional oscillator. This is quantified by the fact that for the quantum state $\rho(t)$ in the oscillator during our simulations, the higher Fock states are minimally occupied, and the canonical commutation relation, ${\rm Tr}\left(\left[\hat{a},\hat{a}^\dagger\right]\rho(t)\right) = 1$ is valid with less than $1\%$ error at all times. As we decrease Hilbert space dimension in our study, the quantum model stops being a valid description of a nonlinear oscillator, but remains a valid description of a qudit, of which many physical examples exist \cite{Clarke:2008aa,Krantz:2019aa}.

Figure \ref{fig:dimsweep} shows results for QRC models ranging from a qubit, $d=2$, to the approximate nonlinear oscillator of the previous subsection, $d=12$. In all cases, we continue to use measurements of the $\hat{X}$ operator as the output, which in the qubit case becomes the Pauli operator $\hat{\sigma}_x$. As Fig.~\ref{fig:dimsweep}(a) shows, the average RMS error depends on Hilbert space dimension in a non-monotonic fashion, while the spread remains roughly consistent. This is further demonstrated in Fig.~\ref{fig:dimsweep}(b), with the RMS error showing a local minimum around $d=5$ for all training set sizes. Focusing on a training set size of 30 in Fig.~\ref{fig:dimsweep}(c), we see that the best performance occurs at $d=5$ and $d=6$, which is almost a factor of 4 better than the worst performance at $d=2$.

The increase in performance as Hilbert space dimension increases is to be expected in QRC, as an increase in Hilbert space dimension implies an increase in the number of degrees of freedom of the reservoir, and thus, its computational power. The nonlinearity of the system dynamics is key in this regard, as it can create non-Gaussian states in the oscillator. Only for such states can the expectation values $E_{nm}$ of Eq.~\eqref{eqn:Enm} for $n,m>2$ be independent of lower order $E_{nm}$, and thus expand the computational space of the reservoir.

The existence of a local minimum, and the saturation of performance for $d\geq8$ is likely explained by a combination of effects. Firstly, due to the finite decay rate $\kappa$, the expectation values $E_{nm}$, or equivalently the Fock state populations, decay exponentially with a rate that is linearly proportional to photon number. As such, while higher Hilbert space dimension offers more computational variables, information is lost from them at a faster rate.

Secondly, even without finite $\kappa$, information can be lost in the higher order $E_{nm}$ by an information scrambling argument. Since we only measure $E_{10} = \left<\hat{X}\right>$, it is possible that information relevant to sine phase estimation is spread into other $E_{nm}$, and is not accessible in $E_{10}$ during the finite times at which we measure. Finally, the ratio of the drive amplitude $\alpha$ to the drive frequency $\omega$ determines how much energy can enter the oscillator, and as higher Fock states are occupied, the nonlinearity $K$ plays a role as well. This further restricts how much of the computational space of the reservoir is accessible.

While we have good heuristic explanations for the weak dependence of QRC performance on Hilbert space dimension, these cannot explain the significant quantum-classical improvement in both average performance and reliability for intermediate training set sizes seen in Fig.~\ref{fig:QRCvCRC}. These improvements persist even when the QRC is a qubit. This is particularly intriguing, since a single qubit is a \emph{classical} system, and we would not have expected any improvement in this case. In the following subsection we delve into this surprising result further, and show how it highlights the likely explanation for the majority of the quantum-classical improvement seen for any Hilbert space dimension.

\subsection{Origin of the Quantum-Classical Improvement}

When we say that a single qubit is a classical system, we mean that there is a local hidden variable model that can completely describe the state space and dynamical evolution of a single qubit. For our purposes, we will consider the Kochen-Specker model \cite{KOCHEN:1967aa}, which maps a density matrix $\rho$ for a single qubit to the three classical variables $(r,\theta,\phi)$ that parameterize a unit-sphere in spherical coordinates. We consider this hidden variable model for a qubit as a reservoir (HVRC) and simulate its performance at the sine phase estimation task. The dynamical evolution of the HVRC is described by a set of transcendental differential equations, so for simplicity we use the results of the quantum simulations for a qubit and convert to the HV model parameters using the relationships
\begin{align}
  \nonumber&\left(x,y,z\right) = \left({\rm Tr}\left[\hat{\sigma}_x\rho\right],{\rm Tr}\left[\hat{\sigma}_y\rho\right],{\rm Tr}\left[\hat{\sigma}_z\rho\right]\right), \\
  \nonumber&\left(r,\theta,\phi\right) = \left(x^2+y^2+z^2,\arccos(\frac{z}{r}),\arctan(\frac{y}{x})\right),
\end{align}
which one may recognize as the familiar coordinate transformation from Cartesian to spherical coordinates.

\begin{figure}[!t]
  \includegraphics[width=0.95\columnwidth]{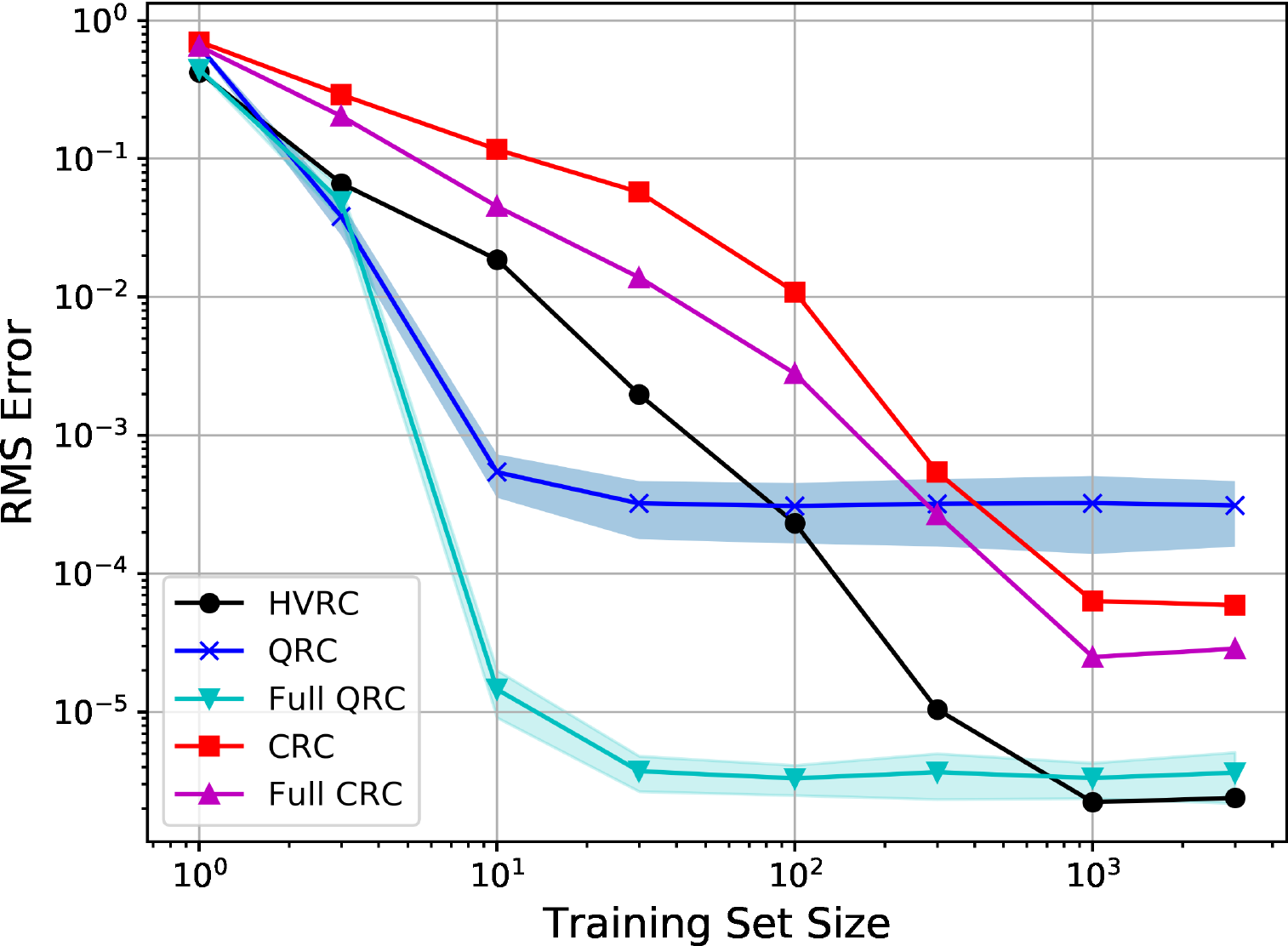}
\caption{{\bf Reservoir performance for qubit and classical models.} As a function of training set size, average root mean square (RMS) error on the test set for a qubit reservoir, either with a single measured expectation value (QRC) or full tomography (full QRC), the hidden variable model for a qubit as a reservoir (HVRC), and classical nonlinear oscillator reservoirs with single quadrature (CRC) or both quadrature (full CRC) measurements. Reservoir parameter set, measurement sampling, and test set size are the same as in Fig.~\ref{fig:QRCvCRC}. One standard deviation for the QRC and full QRC performance is shown in the shaded regions.}
  \label{fig:HV}
\end{figure}

We simulate the performance of the HVRC under the same conditions (parameter sets, training and test sets) as for the QRC and CRC presented previously, and use all three variables as the output. Figure \ref{fig:HV} shows the average performance of the various RC models, and as can be seen, there is considerable performance improvement for the QRC compared to the HVRC at intermediate training set size. There is an even larger improvement for what we call full QRC, which uses full tomographic data on the qubit state (expectation values for all three Pauli operators) as the output, as opposed to QRC which uses only ${\rm Tr}\left[\hat{\sigma}_x\rho\right]$.

We emphasize that all models considered in this subsection describe reservoir dynamics that are purely classical physics. The only difference between the QRC models and the HVRC model is that the HVRC model uses the output variables $(r,\theta,\phi)$, while the QRC models use the output variables $(x,y,z) = (r\sin\theta\cos\phi,r\sin\theta\sin\phi,r\cos\theta)$, which are a nonlinear transformation of the HVRC output. As such, the average performance improvement between QRC and HVRC can only be attributed to this nonlinear transformation. Such a nonlinear transformation cannot be implemented by the linear processing of output data that occurs in the typical operation of a reservoir computer, see Eq.~\eqref{eqn:ResOut}. Nevertheless, previous work has shown that appending nonlinear post-processing to the output of a reservoir can have significant impact on performance \cite{araujo2020}, though one may argue that this is costly to implement.

This nonlinear processing of data occurs intrinsically for the quantum reservoir due to quantum measurement, and is therefore part of QRC itself, being responsible for the generation of output data. As in all RC, standard linear post-processing of the output still occurs, but we essentially get an otherwise costly nonlinear processing stage ``for free'' in QRC. As we have shown, the nonlinearity of quantum measurement is the source of performance improvement between the QRC models and the HVRC for a qubit, from which we infer that it is also the main source of improvement of the qubit QRC compared to the CRC models (see Fig.~\ref{fig:HV}, where full CRC uses both quadrature measurements as output). However, the average HVRC performance is better than that for CRC, which implies it is a better reservoir for the sine phase estimation task. Thus, part of the performance improvement of qubit QRC over CRC is also due to the fact that the HVRC underlying the QRC is better suited for the task at hand.

\begin{figure}[!t]
  \includegraphics[width=0.95\columnwidth]{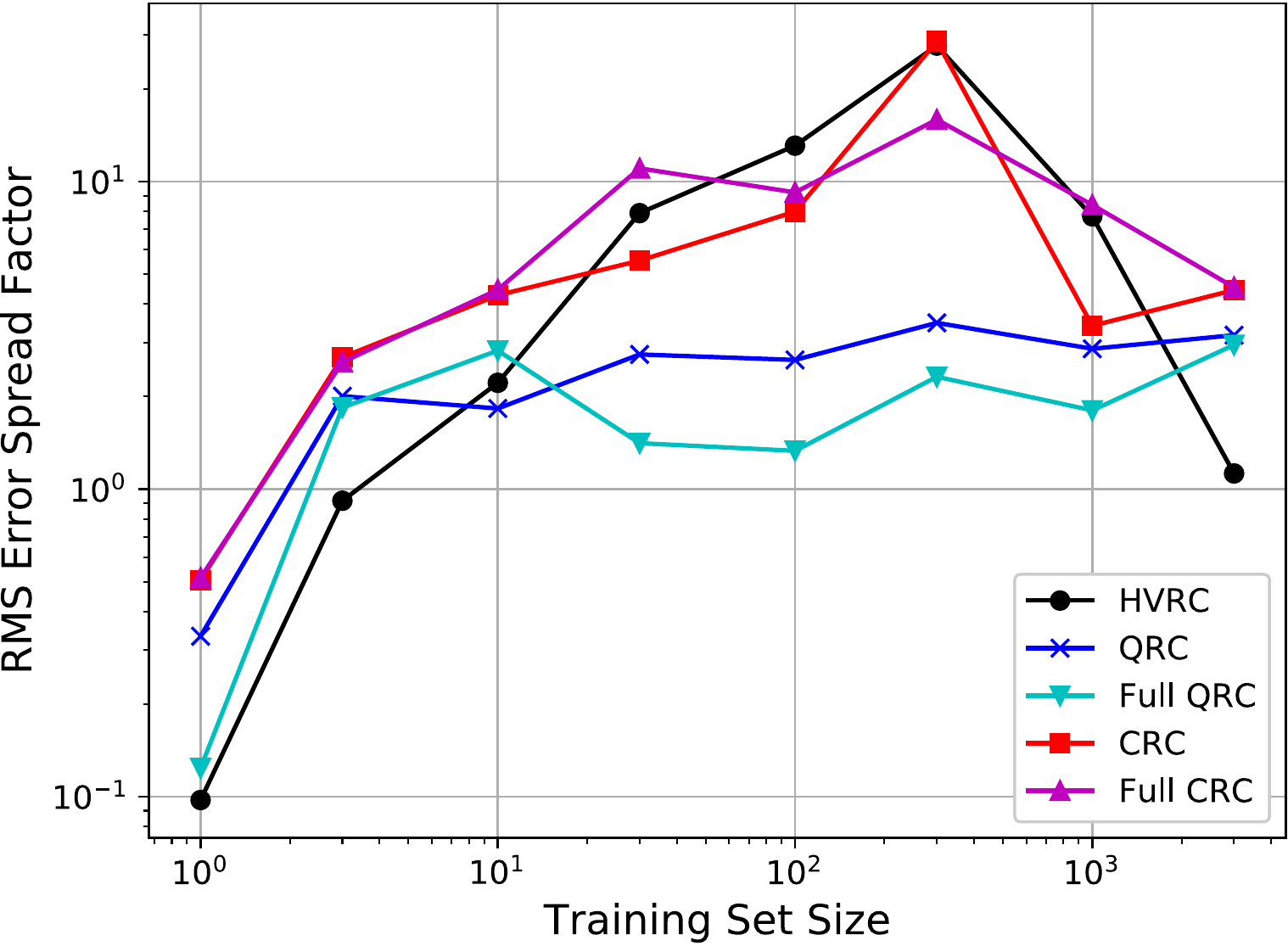}
  \caption{{\bf Performance spread for qubit and classical models.} As a function of training set size, spread in root mean square (RMS) error, cf.~Eq.~\eqref{eqn:spread}, on the test set for a qubit reservoir, either with a single measured expectation value (QRC) or full tomography (full QRC), the hidden variable model for a qubit as a reservoir (HVRC), and classical nonlinear oscillator reservoirs with single quadrature (CRC) or both quadrature (full CRC) measurements. Output sampling and test set size the same as in Figs.~\ref{fig:QRCvCRC} and \ref{fig:HV}.}
  \label{fig:spread}
\end{figure}

While we cannot make definitive statements for QRC beyond $d=2$, as no local hidden variable models exist, the results for the qubit QRC indicate that the quantum-classical performance improvement seen in Fig.~\ref{fig:QRCvCRC} is likely due to the nonlinearity of quantum measurement. Similarly, as shown in Fig.~\ref{fig:spread}, which plots the RMS error spread factor of Eq.~\eqref{eqn:spread}, we see that HVRC has a large spread in performance, similar to CRC. This indicates that the reliability of QRC (at all Hilbert space dimensions) also stems from the nonlinearity of quantum measurement. Thus, due to the improvement in both average performance and reliability, we believe the nonlinear oscillator QRC possesses an advantage over CRC due to the intrinsic nonlinearity of quantum measurement.

\subsection{Input and Output Noise}

As a final consideration for the performance of the nonlinear oscillator RC models, we examine the impact of noise in the both the output data and the input signal. Noise can have a variety of sources, such as thermal noise in the control and readout chains of the device, and will inevitably plague hardware implementations of our models. In particular, for QRC output noise is unavoidable due to the intrinsic uncertainty of quantum measurement. QRC uses the expectation value of an operator as its output variable, which in a single-shot measurement can only be estimated up to quantum uncertainty, often characterized by the standard deviation of the operator expectation value. For the $\hat{X}$ operator we have chosen, this is given by
\begin{align}
  \Delta E_{\rm out} = \sqrt{{\rm Tr}[\rho\hat{X}^2] - {\rm Tr}[\rho\hat{X}]^2}.
\end{align}
For a quantum system with a large (infinite) Hilbert space, such as a nonlinear oscillator, the system can be in a state (e.g.~a coherent state) such that $\abs{E_{01}} \gg \abs{\Delta E_{01}}$. In this case, single-shot measurements give a good estimate of the output variable.

The situation is very different for low dimensional systems. In the extreme case of a qubit, each measurement gives a binary output, and for most quantum states the standard deviation of a Pauli operator measurement is of the same order of magnitude as the expectation value. Thus, it is necessary to average over repeated runs of the same training/test case, to produce can estimate of the output variable, $\tilde{E}_{\rm out}$, with the standard error of this estimate scaling as
\begin{align}
  \Delta \tilde{E}_{\rm out} = \frac{\Delta E_{\rm out}}{\sqrt{R}},
\end{align}
where $R$ is the number of repeated runs. The slow improvement with $\sqrt{R}$ is unfavorable in practice, and importantly, is a feature of all quantum neuromorphic and machine learning protocols that use qubits and rely on real-valued (as opposed to binary) output data. Recently, it was pointed out that one must include the cost of these repetitions when accessing algorithm run time and scaling \cite{Wright:2019aa}.

Returning to the nonlinear oscillator models, it should be clear that we use such models in part to reduce the necessity for repetition, and keep $R$ as small as possible. We artificially introduce output noise by adding a Gaussian random variable with zero mean and standard deviation $\sigma$ to each sampled time point of the output from the reservoir. We do this for the training set, before calculating the weights using Eq.~\eqref{eqn:train}, and for the test set. Due to numerical simulation constraints, we consider a single set of reservoir parameters given by the mean values discussed at the beginning of section \ref{sec:Results}.

\begin{figure}[!t]
  \includegraphics[width=0.95\columnwidth]{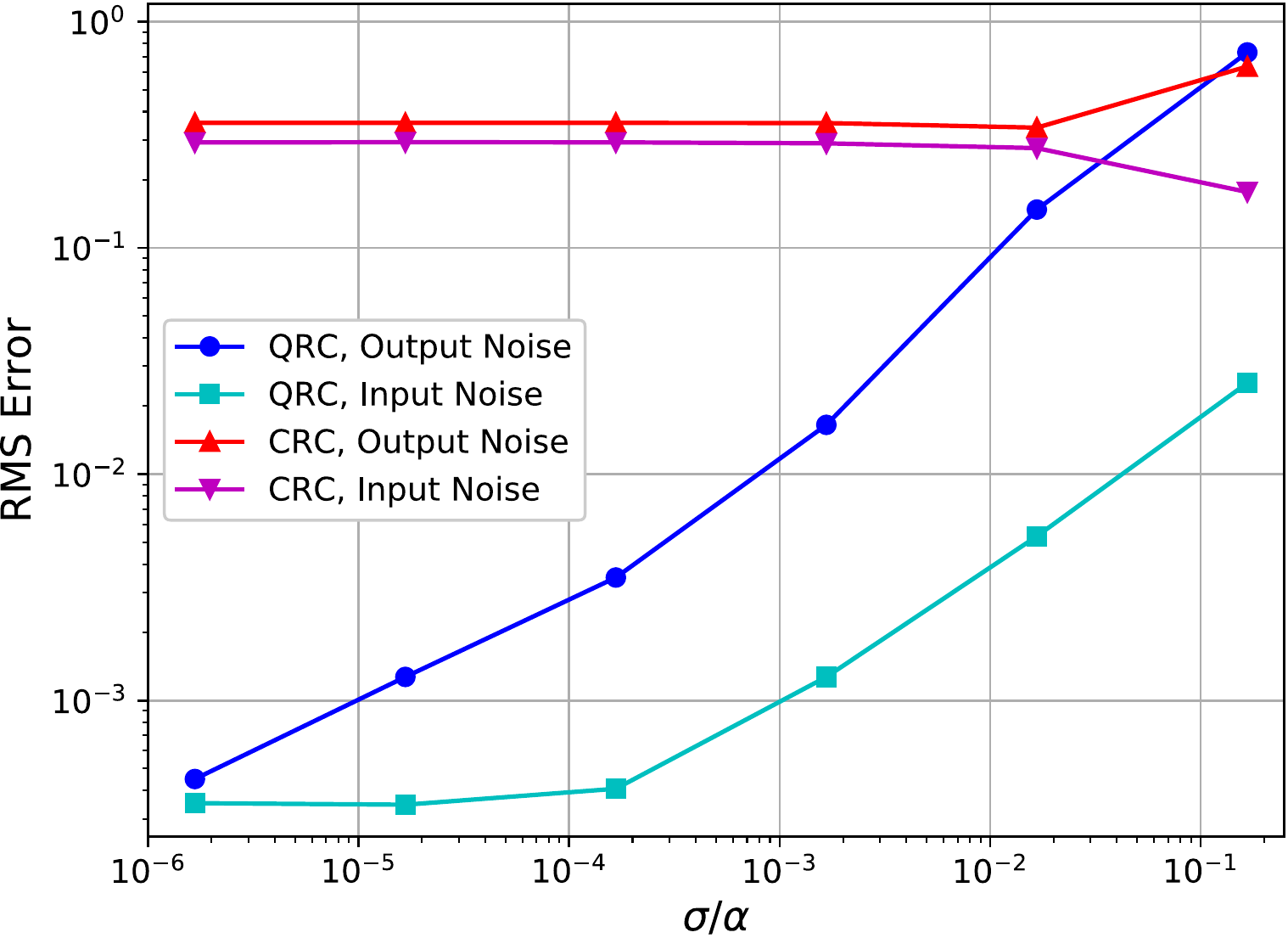}
  \caption{{\bf Quantum and classical reservoir performance with noise.} Root mean square (RMS) error on the test set for the quantum nonlinear oscillator reservoirs (QRC) with Hilbert space dimension $d=12$, and the classical nonlinear oscillator reservoirs (CRC), as a function of noise standard deviation in units of the input drive amplitude ($\sigma/\alpha$) for noise injected in either the output or the input. These results are for a single set of reservoir parameters described in the main text. Output sampling is as in Fig.~\ref{fig:QRCvCRC}, the training set size is 10, and the test set size is 5000 for output noise and 100 for input noise.}
  \label{fig:InNOut}
\end{figure}

The dependence of QRC performance on output noise is shown in Fig.~\ref{fig:InNOut}. As is to be expected, the RMS error increases monotonically as a function of the output noise. We measure the output noise in units of the input signal amplitude $\alpha$, since the output signal expectation value will be a function of $\alpha$. The results of Fig.~\ref{fig:InNOut} indicate that some repetition will be required for high performance in an experimental implementation of this nonlinear oscillator QRC, as it is unlikely that the combination of quantum measurement noise and other classical output noise sources can be brought below $10^{-3}$ (in units of the signal amplitude).

However, numerical simulation resources limit the maximum input signal amplitude we can consider for our reservoirs. It is likely that by increasing input signal amplitude, and therefore output signal amplitude, we can improve the performance of QRC in the presence of output noise, as the output noise would become a diminishing fraction of the output signal. An alternative approach would be to consider nonlinear oscillators with \emph{in situ} parametric driving, such that they act as both a self-amplifier of their output signal, and reduce quantum measurement noise by squeezing. The study of both these routes towards output noise insensitivity in QRC will be the subject of future theoretical and experimental effort.

Now turning to input noise, we artificially inject noise into the input signal to our quantum reservoir. To ensure that our simulations observe causality we cannot use the high-performance adaptive time-step differential equation solvers that are default in {\tt MESolve.jl}, but must switch to a considerably slower fixed time-step solver. We use a time-step $\delta t\ll 1/(50\bar{\kappa}),~1/\bar{\omega}$ that is much smaller than the output sample time-step, the oscillator lifetime, and the input signal oscillation period to ensure the noise closely mimics white noise with no temporal correlations. We add independent Gaussian random noise with zero mean and standard deviation $\sigma = s\sqrt{\delta t}$ at each time-step, such that our simulations closely approximate the effects of Gaussian white noise in the input signal, where $s$ is a scale factor describing the noise spectral amplitude.

Figure \ref{fig:InNOut} also shows the results of our input noise simulations, which as can be seen, are much more favorable than for output noise. In particular, we do not observe an increase in RMS error until $\sigma/\alpha$ reaches a level comparable to the RMS error with no noise, after which the RMS error increases roughly linearly. This indicates that the internal reservoir dynamics do not appear to amplify the effect of input noise, and that training is less disrupted by small amounts of input noise compared to output noise. A detailed understanding of the mechanisms behind these observations is the subject of future study, but we do not expect input noise at experimentally accessible levels to have a limiting impact on QRC for our chosen task.

\section{Conclusion}
\label{sec:conc}

In this work, we have introduced an approach to quantum reservoir computing that uses a single nonlinear oscillator as the reservoir. The computational nodes are formed by the complex amplitudes of the eigendecomposition of the system state, or equivalently, the expectation values of a complete basis of observables. We have demonstrated that this quantum reservoir has improved performance compared to the equivalent classical one at sine phase estimation, both in terms of average estimation error, and in reliability of performance across reservoirs with different internal parameters.

By studying the performance dependence on Hilbert space dimension down to the single qubit level, and comparing to the fully-classical hidden variable theory describing the dynamical evolution of a qubit, we have determined that the main source of the quantum-classical improvement is the nonlinearity inherent to quantum measurement. The impact of this extends beyond reservoir computing, and to our knowledge this has not previously been identified as a beneficial feature intrinsic to all quantum neuromorphic computing approaches. We believe this further divides quantum and classical neuromorphic and machine learning methodologies, strengthens the case for potential (heuristic) quantum advantage, and may shed new light on previous results \cite{Wilson2018}.

With an eye towards experimental implementation, we studied the impact of injected input or output noise on performance, and have found that output noise is the more detrimental. While we have proposed several mitigation strategies, it appears that some repetition of experiments will be necessary to obtain accurate expectation values, as is the case in most proposed implementations of quantum neuromorphic computing or quantum machine learning. Nevertheless, we expect the required number of repetitions to scale more favorably for our continuous variable reservoir than for discrete variable approaches.

The nonlinear oscillator reservoir we have proposed is well suited to an implementation in realistic experiments. In particular, the tools and techniques of circuit quantum electrodynamics (cQED) \cite{Blais2007} that have been developed to build conventional quantum processors provide an attractive realization of the building blocks needed for QRC. cQED experiments revolve around the control of the quantum degrees of freedom of well-isolated nonlinear oscillators (e.g. superconducting qubits), and their basic ingredient, the Josephson junction, affords a high degree of control over circuit parameters \cite{Koch07,Krantz:2019aa}, as well as strong intrinsic nonlinearity. Further, both nonlinear opto- or electromechanical resonators \cite{Sankey:2010aa,Rips2014,Lemonde:2016aa}, and nonlinear photonic systems \cite{Strekalov_2016,Lin:17} are attractive test beds for continuous variable QRC. We expect that experimental implementations of the ideas presented in this work will provide a new area of applications for devices heretofore used in quantum information research.
\\
\acknowledgements

The authors acknowledge insightful discussions on reservoir computing with Daniel Gauthier. This material is based upon work supported by the U.S. Army Research Office under Contract No: W911NF-19-C-0092. Any opinions, findings and conclusions or recommendations expressed in this material are those of the authors and do not necessarily reflect the views of the U.S. Army Research Office.

\bibliography{QRC_Single_Oscillator.bib}

\end{document}